\documentclass[12pt,a4paper,dvips]{article}
\usepackage{epsfig,wrapfig,times,mathptm} 
\renewcommand{\section}[1]{\vspace{6pt} \noindent\mbox{#1} \newline \noindent}
\renewcommand{\subsection}[1]{\vspace{6pt} \noindent\mbox{\underline{#1}} 
\newline \noindent}
\renewcommand{\subsubsection}[1]{\vspace{6pt} \noindent\mbox{\underline{#1}}
\noindent}

\newfont{\sansb}{cmssbx10}
\newfont{\sans}{cmss10}

\begin{document}
{\small OG 4.2.7 \vspace{-24pt}\\}     
{\center \LARGE A SEARCH FOR TEV EMISSION FROM UNIDENTIFIED SOURCES IN
THE GALACTIC PLANE
\vspace{6pt}\\}

J.H. Buckley$^1$, 
P.J. Boyle$^2$,
S.M. Bradbury$^3$, 
A.M. Burdett$^3$, 
J. Buss\'{o}ns Gordo$^2$, 
D.A. Carter-Lewis$^4$, 
M. Catanese$^4$, 
M.F. Cawley$^5$, 
D.J. Fegan$^2$, 
J.P. Finley$^6$, 
J.A. Gaidos$^6$, 
A.M. Hillas$^3$, 
F. Krennrich$^4$, 
R.C. Lamb$^7$, 
R.W. Lessard$^6$, 
C. Masterson$^2$, 
J.E. McEnery$^2$, 
G. Mohanty$^4$, 
J. Quinn$^{1,2}$, 
A.J. Rodgers$^3$, 
H.J. Rose$^3$, 
A.C. Rovero$^8$,
F.W. Samuelson$^4$, 
G.H. Sembroski$^6$, 
R. Srinivasan$^6$, 
T.C. Weekes$^1$, 
and J. Zweerink$^4$ \vspace{6pt}\\
{\it $^1$F. L. Whipple Observatory, Harvard-Smithsonian CfA, P.O. Box 97, Amado, AZ 85645, USA\\
$^2$Dept. of Exp. Physics, University College Dublin, Belfield, Dublin 4, Ireland\\
$^3$Department of Physics, University of Leeds, Leeds, LS2 9JT, Yorkshire, England, UK\\
$^4$Dept. of Physics and Astronomy, Iowa State University, Ames, IA 50011, USA\\
$^5$Physics Department, St. Patrick's College, Maynooth, County Kildare, Ireland\\
$^6$Department of Physics, Purdue University, West Lafayette, IN 47907, USA\\
$^7$Space Radiation Laboratory, California Institute of Technology, Pasadena,
CA 91125, USA\\
$^8$Instituto de Astronom\'{i}a y F\'{i}sca del Espacio, CC 67, Suc.\ 28, (1428) Buenos Aires, Argentina \vspace{-12pt}\\ }
{\center ABSTRACT\\}
The $\sim$70 unidentified sources of the EGRET sky survey may be one of
its most important legacies. The identification of
these sources at other wavelengths is critical to 
understanding their nature. Many have flat spectra out to 10 GeV which, if
extrapolated to TeV energies, would be easily detectable relative to
the steeply falling diffuse background. The Whipple Observatory
$\gamma$-ray telescope has been used to observe a number of these which
were selected based on their
position, intensity and spectrum and in some cases based on a possible
association with a supernova remnant or pulsar.  No significant emission has
been detected from these sources, and upper limits are given.

\setlength{\parindent}{1cm}
\section{INTRODUCTION}
Despite extensive searches for counterparts to the $\sim$30 low-lattitude
unidentified EGRET sources,
the nature of these objects is still largely unknown.
Kaaret \& Cottam (1996) have suggested that the low latitude
unidentified sources show a correlation
with OB associations, the sites of massive star formation. 
Since nearly half of all supernovae occur as the
core collapse of young massive stars which explode into star formation
regions  (e.g., Huang \& Thaddeus 1986) a correlation with the positions
of pulsars and with supernova remnants also follows.  
Kaaret \& Cottam argue that pulsars emit $\gamma$-rays over a
significantly longer lifetime than SNR, so that the number of
$\gamma$-ray pulsars is expected to be
significantly larger than the number of $\gamma$-ray
SNR.  However, Esposito et al.\ (1996), Sturner \& Dermer (1994) and
Sturner, Dermer and Mattox (1996) have presented evidence for associations
of a number of these
objects with SNRs ($\gamma$-Cygni, IC443, W44, Monoceros) 
for which there is no pulsar within the 95\% confidence error
contour (Sturner, Dermer \& Mattox 1996).  Attempts to detect
radio pulsars in the error boxes of EGRET unidentified
sources have been unsuccessful (e.g.,
Nice \& Sayer 1997) and provide some  
constraints on models in which all of the Galactic unidentified
sources are pulsars.  Since EGRET generally lacks the
 spatial resolution to distinguish
the point-like emission from pulsars and AGNs 
from the extended emission expected
to arise in the vicinity of supernova shells, 
variability has been used to distinguish compact sources.  
Dramatic transient sources such as the enigmatic 2CG~135+1
and newly discovered GRO~J1838+04
are difficult to interpret as either arising from AGNs or from pulsars,
and are possibly representatives of a new class of Galactic $\gamma$-ray source
distinct from isolated pulsars (e.g., Tavani et al.\ 1997).  

Detection of these objects at high energies using ground based $\gamma$-ray
detectors would aid in the
identification of these objects in two important ways.
First, the contribution from the diffuse $\gamma$-ray background falls as
a steeper power of energy ($\sim E^{-2.7}$) than the source spectrum
($\sim E^{-2}$)
for many of these objects, implying a smaller effect from uncertainties
in the diffuse background model in determining the position, flux and 
spectra of these sources.  Second, the 0.13$^\circ$ angular resolution of the
Whipple 10m $\gamma$-ray telescope (Lessard and Buckley 1997) provides the
ability to resolve extended sources such as SNRs and offers the potential
to narrow the error box for bright sources.

\section{OBSERVATIONS AND ANALYSIS}
The high-energy $\gamma$-ray telescope (Cawley et al.\ 1990) at
the Whipple
Observatory employs a 10~m diameter optical reflector to image \v
Cerenkov light from air showers onto an array of 109 fast
photomultipliers covering a 3$^\circ$ field of view (FOV).
By making use of the distinctive differences in the angular
distribution of light and orientation of the shower images a
$\gamma$-ray signal can be extracted from the large background of
hadronic showers.

Data are generally taken in a differential mode where each 28~min
ON-source run is followed by a 28~min OFF-source control run which is
offset in right ascension to
ensure that the same range of azimuth and zenith angles are sampled.
While this
cancels the zenith angle dependence of the cosmic-ray rate as well as other
systematic effects in the camera, differences in sky brightness between
the ON-source and OFF-source regions can lead to biases.
For some galactic plane sources, such differences
are substantial due to either diffuse emission from the galactic plane or
bright stars.

Systematic effects arising from such brightness differences
can be largely canceled by the procedure of software padding (Cawley
et al.\ 1983).
This procedure consists of adding noise to all pixels of each event so
that matching PMTs in ON-source and OFF-source runs have identical
noise pulse height spectra.   Only PMT signals
which exceed some multiple of this noise level are included in the subsequent
analysis of the shower images (Punch et al.\ 1993).

The technique used to generate the two-dimensional
plots and upper limits is a simple extension of the standard Whipple
data analysis (Reynolds et al.\ 1993) and is described
in more detail in Lessard and Buckley (1997).
After initial processing of the shower images including pedestal
subtraction, gain correction and image cleaning (e.g., Punch et al.\ 1993)
individual \v Cerenkov shower images are subjected to a moment analysis
to determine a set of parameters that characterize the roughly elliptical
images.  Each point on a two dimensional grid covering the
3$^\circ$ FOV is considered as the potential
source position.  For each event,
the RMS width and length, centroid position,
orientation, ellipticity 
and the skew of the shower image are calculated about this
point of origin and tested for consistency with the parameter values
expected for a $\gamma$-ray event coming
from the corresponding direction in the sky.
For each grid point the number of candidate events ON-source and OFF-source
are calculated, and the significance of the excess, $S_\lambda$, is
derived using the likelihood ratio method of Li \& Ma (1983).  In the
two-dimensional plots presented in Figures 1, the gray-scale indicates
the number of excess counts (candidate $\gamma$-rays) consistent with
each grid point and the contours shown correspond to
the statistic $S_\lambda$ in steps of 1.    
Note that due to the large number of trials
associated with the 30$\times$30 grid, approximately one $S_\lambda$=3
excess is expected for each two-dimensional plot.

While it is desirable to have one control (OFF-source) run for each
run ON-source, for some of the data presented
here the number of exposures taken OFF-source is less than
the number of ON-source runs.  In this case, the
background level is determined by normalizing the OFF-source data to the
ON-source data in 
a 0.25$^\circ$ band around the perimeter of the field of view.  The resulting
normalization factor $\alpha$ enters into the calculation of the significance
and the upper limit
following the procedure of Li and Ma (1983). 
For the sources J0542+26, J0635+0521, and J1825-1307,
little or no OFF-source data was taken and
a background template was formed using contemporaneous control data taken
for other sources.  This results in an additional systematic error for
these sources.

In calculating upper limits, we are testing the hypothesis that the 
emission is coming from a point source within the EGRET error box.
Upper limits calculated for the extended regions corresponding to
the
IC443, W44 and $\gamma$-Cygni SNRs (for 2EG J0618+2234,
2EG J1857+0118, and 2EG J2020+4026 respectively) are reported elsewhere
(Buckley et al.\ 1997).
99.9\% confidence-level upper limits are calculated for each grid point
lying within the EGRET
95\% confidence interval using the method of Helene (1983) and accounting
for the declining
$\gamma$-ray detection efficiency
away from the camera center (Lessard \& Buckley 1997).
The maximum upper limits for each error box are shown in Table 1.

\section{RESULTS AND CONCLUSIONS}
\label{results}
Table~1 shows preliminary
Whipple upper limits for a number of unidentified sources together
with the extrapolated EGRET flux derived from the EGRET spectrum.
Fluxes and spectral indices are from Fierro et al.\ (1996).
In addition to sources from the first (Fichtel et al.\ 1994)
or second EGRET catalogs (Thompson et al.\ 1995), we also
include the source J0749+17 from the initial list of unidentified
sources by Hartman et al.\ (1992).   
This object was not included in the
first EGRET catalog  because of its low significance
($<4\sigma$),
but is of interest since it prompted the radio pulsar search by
Lundgren, Zepka and Cordes (1995) that led to the discovery of the
binary millisecond pulsar PSR~0751+1807.   Also on our list is
the source J0542+26 which was on the list of high confidence unidentified
sources in the first but not the second EGRET catalog.  This source
has a 158~arcmin error radius which could not be shown in Figure 1.
This source is of interest since it is coincident with the position of
the old, nearby (0.8-1.4kpc, Kundu et al.\ 1980) SNR 
S~147 as pointed out by Sturner and Dermer (1994).

These data were taken over the period December 1994 to May 1997.
Two-dimensional plots for these sources are shown in Figure 1 excluding
results for J0618+2234, J1857+0118 and J2020+4026 which are
shown elsewhere (Buckley et al.\ 1997). Upper limits
are at energies above 400~GeV unless otherwise indicated.  2EG~1746-2852 
transits at an elevation of $<$30$^\circ$ resulting in an increase
of the effective area and energy threshold by a factor of approximately
5.0 compared with observations at the zenith (Krennrich et al.\ 1997).
While 2EG~J1746-2852 shows a small (2.5$\sigma$) excess at the position of
Sgr~A$^*$ and within the EGRET error box, this excess is not considered
significant given the additional systematic errors present for galactic
plane sources.  2EG~J0241+6119 shows a similar excess near the
position of 2CG135+01 and within the EGRET error box.  The
excess in J0542+26 lies outside and to the south of the radio shell
of S147 and approximately 0.5$^\circ$ away from the X-ray binary
4U0535+262, too far to make an association.  The other sources
show no significant emission within the EGRET error boxes.
Further deep observations with the GRANITE-III high resolution camera
should provide better sensitivity given the extended FOV and
finer pixelization, and correlations with data taken at other wavelengths
should improve chances for detecting variable unidentified sources. 

{\footnotesize
\begin{table*}
\caption{Results of Observations. \label{table:results}}
\footnotesize
\begin{tabular}{lrrlrrrrr} \hline\hline
  & & & & \multicolumn{2}{c}{EGRET ($>100$~MeV)} & 
\multicolumn{1}{c}{Prediction}
 & \multicolumn{2}{c}{Whipple ($>400$~GeV)}
\\ \cline{5-6} \cline{8-9}
  & \multicolumn{2}{c}{Position} & \multicolumn{1}{c}{Nearby}
 & \multicolumn{1}{c}{Flux $(10^{-8}$} & \multicolumn{1}{c}{Spectral} &
\multicolumn{1}{c}{$(>400$GeV$)$} & \multicolumn{1}{c}{Exposure}
 & \multicolumn{1}{c}{Flux Limit} \\ \cline{2-3}
\multicolumn{1}{c}{Name} & \multicolumn{1}{c}{l} & \multicolumn{1}{c}{b}
  & \multicolumn{1}{c}{Objects}  & 
\multicolumn{1}{c}{cm$^{-2}$s$^{-1}$)} & \multicolumn{1}{c}{Index}  
 & \multicolumn{3}{c}{\footnotesize($10^{-11}$cm$^{-2}$s$^{-1}$) (min) ($10^{-11}$cm$^{-2}$s$^{-1}$)} \\ 
\hline
J0241+6119 & 135.74 & 1.22 & 2CG135+01 & 87.0$\pm$6.7 & -2.2$\pm$0.1
& 4.14 & 972 & 1.02 \\
J0542+26   & 181.92 & -2.00 & S147 & 17.6$\pm$3.5 & -3.3$\pm$0.5
&  &  &   \\
J0545+3943 & 170.79 & 5.65 &           & 12.7$\pm$2.8 & -3.0$\pm$0.3
& & 108 & 6.72 \\
J0618+2234 & 189.13 & 3.19 & IC443 & 45.7$\pm$3.8 & -1.8$\pm$0.1
& 70.8 & 1188 & 0.911 \\
J0635+0521 & 206.30 & -1.20 & Monoceros & 24.5$\pm$4.1 & -2.4$\pm$0.3
& 3.09 & 108 & 5.59 \\
J0749+17   &        &       & PSR 0751+1807 &         &
&  & 486 & 0.813 \\
J1746-2852 & 0.17   & -0.15 & Sgr A$^*$ & 110.9$\pm$9.4 & 
&  & 270 & 0.45$^\dagger$ \\
J1825-1307 & 18.38  & -0.43 & PSR B1823-13 & 84.0$\pm$7.9 & -2.3$\pm$0.1
& 1.48 & 702 & 1.55 \\
J1857+0118 & 34.80  & -0.76 &  W44  & 52.1$\pm$8.7 & -1.9$\pm$0.2
& 29.9& 351 & 2.79 \\
           &        &       & PSR~B1853+01 &          & & & & \\
J2020+4026 & 78.12 & 2.23 & $\gamma$-Cygni SNR &122.9$\pm$6.8& -2.0$\pm$0.1
& 33.8 & 513 & 0.990 \\
\hline
\end{tabular}
\noindent $^\dagger$ Integral flux above 2.0~TeV.
\vspace*{-0.2in}
\end{table*}
}

\begin{figure}[t]
\label{plots}
\vspace*{-1.0in}
\hspace*{-0.5in}\epsfig{file=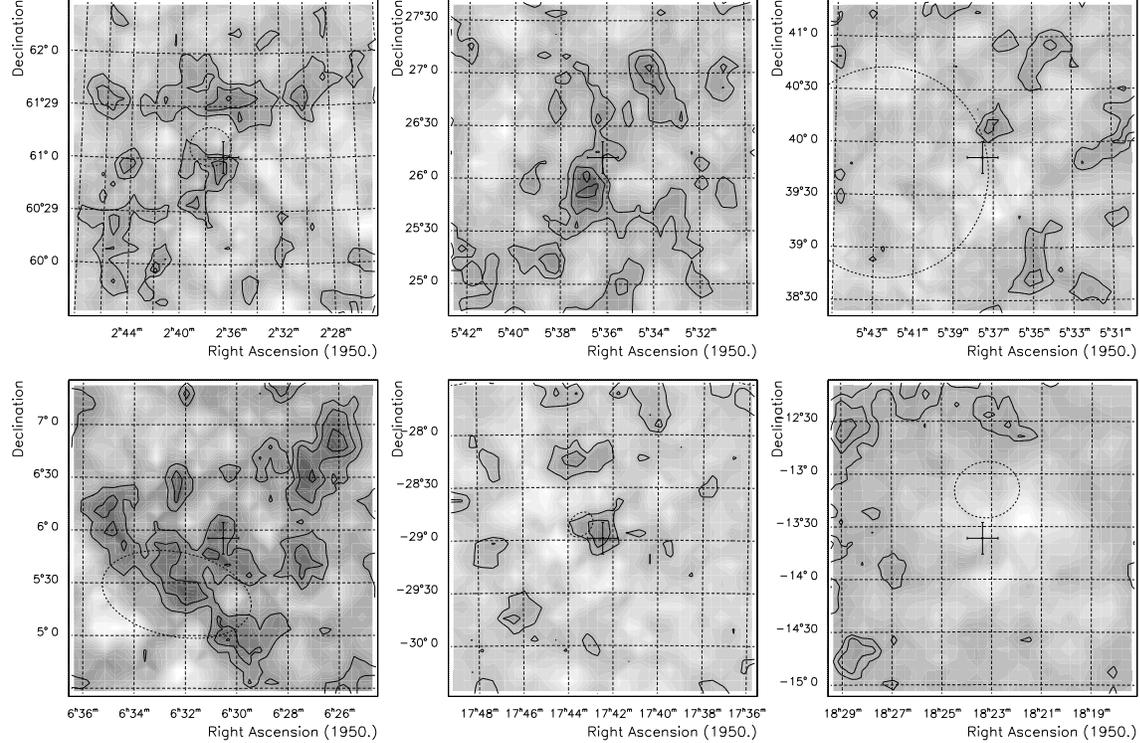,height=10.0in,angle=0.}
\vspace*{-5.3in}
\caption{\it Plots of two-dimensional distributions of candidate
gamma-ray events and contours of $S_\lambda$
for (a) J0241+6119 (cross at the position of 2CG135+01), (b) J0542+26,
 (c) J0545+3943, 
(d) J0635+0521, 
(e) J1746-2852 (cross at Sgr~A$^*$)
and (f) J1825+2234 (cross at the position of PSR B1823-13).
Dotted contours
show the elliptical fit to the EGRET 95\% confidence intervals.
.}
\vspace*{-0.1in}
\end{figure}

\section{ACKNOWLEDGEMENTS}
This research is supported by grants from the U. S. Department of
Energy and NASA, by PPARC in the UK and by Forbairt in Ireland.

\section{REFERENCES}
\setlength{\parindent}{-5mm}
\begin{list}{}{\topsep 0pt \partopsep 0pt \itemsep 0pt \leftmargin 5mm
\parsep 0pt \itemindent -5mm}
\vspace{-15pt}

\item Buckley, J.H., et al., in preparation (1997).

\item Cawley, M.F, et al., {\it Exper.\ Astr.}, 1, 173 (1990).

\item Esposito, J.A., et al., {\it ApJ}, 461, 820 (1996).

\item Fichtel, C.E., et al., {\it ApJS}, 94, 551 (1994).

\item Fierro, J., Ph.D.\ Dissertation, Stanford University (1996).

\item Hartman, R.C., et al., {\it BAAS}, 24, 1155 (1992).

\item Helene, O., {\it Nucl. Instr. Meth.}, 212, 319 (1983).

\item Huang, Y.-L., \& Thaddeus, P., {\it ApJ}, 309, 804 (1986).

\item Kaaret, P., \& Cottam, J., {\it ApJ}, 462, L35 (1996).

\item Krennrich, F., et al., {\it ApJ}, in press (1997).

\item Kundu, Angerhofer, P.E., F\"urst, E., \& Hirth, W., {\it A\&{A}}, 92, 225 (1980). 

\item Lessard, R.W., \& Buckley, J.H., in preparation (1997).

\item Li, T.-P., \& Ma, Y.-Q., ApJ, 272, 317 (1983).

\item Lundgren, S.C., Zepka, A.F., \& Cordes, J.M., {\it ApJ}, 453, 419 (1995).

\item Nice, D.J., \& Sayer, R.W., {\it ApJ}, 476, 261 (1997).

\item Punch, M., Ph.D.\ thesis, National University of Ireland (1993).

\item Reynolds, P. T., et al., {\it ApJ}, 404, 206 (1993).

\item Sturner, J.A., Dermer, C.D., {\it A\&{A}}, 293, L17 (1994).

\item Tavani, M., et al., {\it ApJ}, 479, 109 (1997).

\item Thompson, D.J., et al., {\it ApJS}, 101, 259 (1995).
\end{list}

\end{document}